\documentstyle[aps,12pt]{revtex}

\def\beq{\begin{equation}}
\def\eeq{\end{equation}}
\def\pa{\partial}
\def\qb{{\bar q}}
\def\la{\langle}
\def\ra{\rangle}

\def\pu{p^\mu}

\def\ku{k_\mu}
\def\varp{\varepsilon}
\def\bea{\arraycolsep .1em \begin{eqnarray}}
\def\eea{\end{eqnarray}}

\def\vp{{\bf p}}

\def\vk{{\bf k}}

\let\de=\delta
\let\eps=\epsilon

\def\eq#1{Eq.(\ref{#1})}
\def\refr#1{$^{\cite{#1}}$}

\def\s0#1#2{\mbox{\small{$ \frac{#1}{#2} $}}}
\def\0#1#2{\frac{#1}{#2}}

\begin{document}
{\begin{center}

{\normalsize\begin{flushright}
{\tt \underline{hep-ph/9907369}}\\[12ex] 
\end{flushright}}

\mbox{\large \bf
{Nonlinear Permittivity Including Non-Abelian }}\\[3ex]
\mbox{\large \bf {Self-interaction of Plasmons in Quark-Gluon Plasma}}\footnote{
Supported by the National Natural Science Foundation of China under Grant No.19775017}
\\[6ex]
{Chen Ji-sheng
\footnote{E-Mail: Chenjs@iopp.ccnu.edu.cn}
$^{1}$
~~~ Zheng Xiao-ping$^2$
and
Li Jia-rong$^1$
\\[4ex]
{\it $^1$
Institute of Particle Physics,
Huazhong Normal University,
Wuhan 430079,P.R.China.
\\
$^2$Department of Physics,
Zhongshan University,
Guangzhou 510275,P.R.China.
}\\[6ex]
\begin{minipage}{14cm}{\small }
~~~~By decomposing the distribution functions and color field to regular and fluctuation parts,
the solution
 of the semi-classical kinetic equations of quark-gluon plasma is analyzed.
Through expanding the kinetic equations  of the fluctuation parts to third order, the
nonlinear permittivity including the self-interaction of gauge field is obtained and
a rough numerical estimate is given out for the important $\vk =0$ modes of the pure gluon plasma.
\end{minipage}}
\end{center}}\vskip 0.25cm
{\bf PACS}:12.38.Mh, 51.10.+y, 77.22.Ch
\newpage
\par
During recent years, the energy changes of high energy partons
traversing quark-gluon plasma(QGP) have been the subject of intensive
interest\refr{wangxn4}.
To
study the behavior of the high energy partons through QGP, one must
constitute a kind of response theory for QGP to external current.
 The
color
permittivity, characteristic quantity of QGP media, must be known.
In the frame of kinetic theory,
the Abelian-like color permittivity of QGP under the linear
approximation has been given out long ago\refr{Heinz}. 
Just as Ref.\cite{markov} pointed out, the self-coupling terms
of color field have not been included in the calculation of relevant quantities.
Ref.\cite{zheng} gave out the nonlinear color permittivity of QGP
beyond the linear approximation,
in which the $SU(N_c)$ color algebra has been ensured
in the iteration process and, to some extent, the non-Abelian characteristic of QGP
has been reflected. However,
there
still exists the problem pointed out by Ref.\cite{markov}.
\par
In this letter, the color permittivity
of QGP including the self-coupling contribution of the color field is calculated with the kinetic theory.
Recently, a few works have been done by separating the distribution
functions
and color field into regular and fluctuation parts, with which the semi-classical kinetic
equations are converted to the forms which can be solved conveniently\refr{markov,zheng,litim}.
The quantities in the kinetic equations
 of QGP, such as the quark or antiquark distribution function $f_q(x,p)$ or 
$f_\qb (x,p)$, gluon distribution function $G(x,p)$
, field $A(x)$ and induced color current $j$ in QGP, can be decomposed as
\bea
f_q=f^R_q+f_q^T,~~~A_\mu=A_\mu
^R+A^T_\mu,~~~j=j^T+j^R,
\eea
where the index $R$ represents the regular parts of corresponding quantities, i.e.,
$f^R=\langle f \rangle$,  
with index $T$ the stochastic fluctuation parts.
$\la \cdots \ra $ represents taking average over a statistical ensemble.
Still further, the distribution functions $f_q^T$,..., can be expanded as the series of small
fluctuation $A^T$, for example
\bea\label{expansion}
f_q^T=\sum _{n=1}^{\infty}f_q^{T(n)}.
\eea
\par
With this kind of decomposition, one can obtain the kinetic equations for the fluctuation parts of quark, antiquark and gluon distribution functions
from the semi-classical kinetic equations of QGP, respectively$^{\cite{Heinz}}$. 
If considering only the fluctuations
around the equilibrium state, 
$f_q^R ( f_{\bar q}^R )$, $G^R$
 can be chosen as Fermi-Dirac and Bose-Einstein equilibrium distribution
functions, respectively.
One can set $A^R=0$ and need only to consider the evolution equation of each fluctuation quantity.
For example, the
fluctuation part of quark distribution function satisfies
\bea\label{original}
\pu &&\pa _\mu f^T_q=\nonumber\\
&&ig ([A^T_\mu,f^T_q]-\la [A^T_\mu,f^T_q]\ra)
-\012 g\pu \{F^T_{\mu\nu L} ,\0 {\pa f^R _q}{\pa p_\nu}\}
 +\012 ig^2p^\mu  \{[A^T_\mu,A^T_\nu]-\la [A^T_\mu,A^T_\nu]\ra ,
\0 {\pa f^R_q}{\pa p_\nu}\}
\nonumber\\
&&-\012 g\pu (\{F^T_{\mu\nu L},\0 {\pa f^T_q}{\pa p_\nu} \}
-\la \{F^T_{\mu\nu L},\0 {\pa f^T_q}{\pa p_\nu}  \} \ra )
+\012 ig^2p^\mu  (\{[A^T_\mu,A^T_\nu],\0 {\pa f^T_q}{\pa p_\nu}\}
-\la \{[A^T_\mu,A^T_\nu],\0 {\pa f^T_q}{\pa p_\nu}\}\ra  ).\nonumber\\
\eea
The kinetic equations of the fluctuation parts of the antiquark
 and gluon distribution functions are similar to
\eq{original} except
the opposite signs of the terms related to $\{...,...\}$ for antiquarks;
 $f$, $A$ and $F_{\mu\nu }$ are replaced by $G$, ${\cal A}$ and ${\cal F}_{\mu\nu }$
 for gluons, respectively.
 $F_{\mu\nu }=F^a_{\mu\nu }t_a$,
${\cal F}_{\mu\nu }=F^a_{\mu\nu }T_a$, with $A=A^at_a$ and ${\cal A}=A^aT_a$.
The $t_a$ and $T_a$ are the corresponding generators of $SU(N_c)$ in fundamental and adjoint representation, respectively.
In the following discussion
we will only give out explicitly the relevant equations for $f_q^T$, but not for
$f_\qb^T$ and $G^T$ unless specialized otherwise.
In \eq{original}, the index $L$ denotes the linear term of $F_{\mu \nu}$ with respect to
$A^T_\mu$.
\par
The fluctuation part  $A^T$ of color field $A$ obeys the following Yang-Mills
equations
\bea\label{yang}
\pa _\mu &&F^{T\mu\nu}_L
=- j^{T\nu}
 +ig \pa _\mu
([A^{T\mu},A^{T\nu}]-\la [A^{T\mu},A^{T\nu}]\ra )+\nonumber\\&&
+ig ([A^T_\mu , F^{T\mu \nu }_L ]-\la [A^T_\mu , F^{T\mu \nu}_L ] \ra)
+g^2([A^T_\mu ,[A^{T\mu},A^{T\nu} ]]-\la [A^T_\mu ,[A^{T\mu},A^{T\nu}]] \ra
),
\eea
and $j^{T\nu }$ can be expanded according to \eq{expansion}, 
with
\bea\label{current}
j^{T(n)\nu}=gt^a\int\0 {d^3p}{(2\pi )^3p^0}p^\nu [{\em tr} t^a(f^{T(n)}_q-f^{T(n)}_{\bar q})+{\em Tr}
(T^aG^{T(n)})],
\eea
being the n-th order color current. 
\par
It is convenient to work in  the temporal axis gauge, i.e., $A_0=0$ and in the momentum space.
The index $T$ of field will be omitted as
well, if without confusion.
By substituting the n-th order distribution function $f^{(n)}(k,p),\cdots$ obtained from the n-th order kinetic equations into \eq{current}, one can obtain $j^{T(n)}(k)$. 
If expanding $j^{T}(k)$ up only to the first order current $j^{T(1)}(k)$, one can recover the
Abelian-like permittivity $\varepsilon _L$.  
It can be confirmed that
the second order color current does not contribute to the nonlinear
permittivity\refr{zheng,liboff}. So, to obtain the nonlinear permittivity including the
self-interaction of gauge field, one needs to  analyze the third order equations directly.
\par
The following shorthands 
\bea
\sum _{k=k_1+k_2}=\int \de (k-k_1-k_2)
dk_1dk_2\0{1}{(2\pi)^4},
~~~ ~~{\cal N}_{eq}=\012\left (f^{(0)}_q(p^0)+f^{(0)}_\qb (p^0)\right )+N_cG^{(0)}(p^0).
\nonumber\\
\sum _{k=k_1+k_2+k_3}=\int \de (k-k_1-k_2-k_3)
dk_1dk_2dk_3 \0{1}{(2\pi)^8},~~~~~
\chi^{\alpha \beta }(k,p)=(p\cdot
k)g^{\alpha \beta}-p^\alpha k^\beta .
\eea 
are introduced for notational simplicity.
The third order kinetic equation is as following
\bea\label{thirdq}
-&&i\pu \ku  f_q^{T(3)}(k,p)=\nonumber\\&&ig p^i  
\sum _{k=k_1+k_2}\left [([A_i(k_1), f_q^{T(2)}(k_2,p)]-\la [A_i(k_1),
f_q^{T(2)}(k_2,p)] \ra
)+\right. \nonumber\\&&\left. +
\012 ig \sum _{k=k_1+k_2}\chi ^{i\lambda }(k_1,p)\0{\pa }{\pa p^\lambda }
 \left (\{A_i(k_1), f_q^{T(2)}(k_2,p)
\}
-\la \{A_i(k_1), f_q^{T(2)}(k_2,p)\}\ra\right )+\right .\nonumber\\&& +\left .
\012 ig^2 \sum _{k=k_1+k_2+k_3}p^i\0{\pa }{\pa p_j}\left (
\{ [A_i(k_1),A_j(k_2)],f_q^{T(1)}(k_3,p\}-\la
\{[A_i(k_1),A_j(k_2)],f_q^{T(1)}(k_3,p)\}\ra \right
)\right ].\nonumber\\
\eea
Considering
the soft excitation carrying momentum
$k\sim gT$ and
the momentum of particles $p\sim T$ in QGP\refr{Blaizot,bodeker},
by
 substituting the third order distribution functions (\ref{thirdq}) into \eq{current}, and keeping only the
leading order in $g$, one may identify
after some lengthy calculation
\bea\label{thcu}
&&j^{T(3)al}(k)\approx\nonumber\\ &&
\Sigma_{k,k_1,k_2,k_3}^{(I)ijkl}
(A_i^b(k_3)A_j^d(k_1)A^e_k(k_2)-A_i^b(k_3)\la
A_j^d(k_1)A^e_k(k_2)\ra -\la A_i^b(k_3)A_j^d(k_1)A^e_k(k_2)\ra ),
\eea
where
\bea
\Sigma
_{k,k_1,k_2,k_3}^{(I)ijkl}=-g^4\sum _{k=k_1+k_2+k_3}f^{abc}f^{cde}\int \0{d^3p}{(2\pi )^3p^0}
\0{p^ip^jp^kp^l}{p\cdot k +ip^0\epsilon }\0{1}{p\cdot (k_1+k_2)+ip^0\epsilon
}\0{\omega _2\pa _p^0{\cal N}_{eq}}{p\cdot k_2 +ip^0\epsilon}.\nonumber\\
\eea
\par
Now turn to the calculation of the non-Abelian permittivity from the mean
field equation.
By expanding $j^{T}(k)$ up to third
order,
substituting $j^{T(2)i}(k)$ and $j^{T(3)i}(k)$ leading order in $g$ into 
\eq{yang} in temporal axis gauge and in momentum space,
 one can identify
\bea\label{third}&&
-\omega ^2 A(k) +{\bf j}^{T(1)}\cdot \0{\bf k}{K}\nonumber\\
\approx &&-g^2t^a\sum
_{k=k_1+k_2+k_3}\left \{g^2\int
\0{d^3p}{(2\pi
)^3p^0}f^{abc}f^{cde}\0{1}{p\cdot k +ip^0\epsilon}\0{1}{p\cdot (k_1+k_2)+ip^0\epsilon }
\right.\nonumber\\&&\left .
\times\0{{\bf p}\cdot {\bf k}}{K}\0{\vp \cdot \vk _1}{K_1}\0{\vp \cdot \vk _2}{K_2}\0{\vp \cdot \vk _3}{K_3}
\0{\omega _2\pa _p^0{\cal N}_{eq}}{p\cdot k_2 +ip^0\epsilon}
\right .\nonumber\\
&&\left.\times\left (A^b(k_3)A^d(k_1)A^e(k_2)-A^b(k_3)\la A^d(k_1)A^e(k_2)
\ra-\la A^b(k_3)A^d(k_1)A^e(k_2)\ra \right )\right .\nonumber\\
&&\left.-\0{{\bf k}_1\cdot {\bf k}_2}{K_1K_2}\0{{\bf k}\cdot{\bf
k}_3}{KK_3}f^{abc}f^{cde}\left [A^b(k_3)A^d(k_1)A^e(k_2)-\la
A^b(k_3)A^d(k_1)A^e(k_2)\ra  
\right ]\right \}.
\eea
The first term on the right-hand side of above equation with the factor in big parentheses
corresponds to the interaction between the particles and the secondary waves resulting from
the nonlinear interactions of the eigenwaves in QGP.
It should be stressed that the second term on the right-hand side of \eq{third} with the factor
in square brackets is the
plasmon self-interaction term in QGP, which is characteristic of the QCD
plasma and different from the QED.
It is well known that the solution of the Yang-Mills equation in QCD vacuum is
very difficult because of the self-coupling term.
Just as Ref.\cite{markov} pointed out that,
in previous works, this term reflecting the non-Abelian characteristics
has been discarded.
However, in hot dense
medium where the phase of the excitation of field is random, one can take
average over phase in the process of calculation\refr{stinko} and make it possible
to solve the mean field equation (\ref{yang}) or
(\ref{third}).
\par
Multiplying \eq{third} with $A^g(k')$, then taking  average with respect to the random
 phase,
one can identify\refr{stinko}
\bea\label{nper}
\left (\varp ^L \de _{ad}+ \varp ^{NL}_{ad} \right ) \la dg\ra =0,
\eea
with the notation $\la dg\ra=\la A^dA^g\ra _k $ representing the correlation strength of color field.
The nonlinear permittivity defined in \eq{nper} depends on the correlation strength
\bea
\label{nper3}
\varp ^{NL}_{ad}\equiv \varp ^{S}_{ad}&&+\varp ^{R}_{ad},\\
\varp ^{S}_{ad}=&&-\0{g^2}{\omega ^2} \int\left [ f^{abc} f^{cfd}\la bf\ra _{k_1}+ \left (\0{\vk
\cdot \vk _1}{KK_1}\right )^2 (f^{abc}f^{cde}\la be\ra_{k_1} +
f^{adc}f^{cfe}\la fe \ra _{k_1} )\right ]\0{dk_1}{(2\pi )^4},
\nonumber\\
\varp ^{R}_{ad}=&&
-\0{g^4}{\omega ^2}\int\0{d^3p\pa _p^0 {\cal N}_{eq}}{(2\pi )^3p^0}  f^{abc}\left ( \0{ \vp\cdot \vk _1
}{K_1}
\right )^2\left ( \0{ \vp\cdot \vk 
}{K}
\right )^2\0{1}{p\cdot k+ip^0\eps }\nonumber\\
&&\times\0{1}{p\cdot (k-k_1)+ip^0\eps }\left
(\0{\omega }{p\cdot k+ip^0\eps }f^{cfd}\la bf \ra
_{k_1}+\0{\omega _1}{p\cdot k_1+ip^0\eps }f^{cde}\la be \ra_{k_1} \right )
\0{dk_1}{(2\pi )^4}.\nonumber
\eea
The $\varp ^{S}_{ad}$ comes from the self-interaction of plasmons, while
the $\varp ^{R}_{ad}$ represents the contribution from the
the interactions between the plasma
particles and the beating of two timelike eigenwaves.
The factors  ${(p\cdot k
+ip^0\eps )}^{-1}$, ${(p\cdot k _1 +ip^0\eps )}^{-1}$, ${(p\cdot (k-k_1)
+ip^0\eps )}^{-1}$ in $\varp ^R_{ad}$ have different contributions to
the integrals. The two former factors  have only real contributions 
because the eigenwaves in QGP are always timelike\refr{Heinz}, while the latter has
both real and imaginary contributions to $\varp ^R_{ad}$ because the beating
of two timelike eigenwaves can be spacelike 
by noticing the crucial relation \refr{zhang}
\bea
\0{1}{p\cdot (k-k_1) +ip^0\eps }=P\0{1}{p\cdot (k-k_1)}-i\0{\pi}{p^0}
\de (\omega _\vk - \omega _{\vk _1}-\0{\bf p}{p^0}
 \cdot (\vk -\vk _1))
.\eea
\par
From \eq{nper3}, one can conclude that the nonlinear permittivity $\varp ^{NL}_{ad}$ is a matrix in color
space and depends on the correlation $\la A^bA^c \ra _k $.
A rough value estimate can make us
see the contributions of $\varp ^S_{ad}$ and $\varp ^R_{ad}$ to the color permittivity more clearly.
One can easily calculate the
diagonal elements that is considering  only the correlation of the same
color by using 
\bea
\la bc\ra _k=-\0 {\pi }{\omega ^2}(\de
(\omega -\omega _\vk )+\de (\omega +\omega _\vk ))I_\vk \de _{bc},
\eea
where $I_\vk$ characterizes the total intensity of the fluctuating oscillation with frequency $\omega _\vk $.
Furthermore, if considering the fluctuation of thermal level, i.e., one can take $I_\vk=4\pi T$.
By taking the upper limit of $K_1$ to be $gT$ while the
lower limit $g^2T$,
the real part of the diagonal elements of $\varp ^{NL}_{ad}$ can be easily
obtained for the important $\vk =0$ modes of the
pure gluon plasma occasion
\bea\label{final}
Re(\varp ^{NL}_{ad})=\delta ^{ad}(\varp ^S+Re\varp ^R)
\approx 
-\delta ^{ad}(0.976g+1.913g).
\eea
\par
Summarizing,
the non-Abelian permittivity (\ref{nper3}) including the self-interaction of gauge field is given out for the first time.
It relies on the concrete modes $\omega _\vk$ and the correlation of field as in the electro-magnetic plasma. However,
It is truly non-Abelian, because of the presences of color indices and the $SU(N_c)$ structure constants.
The motivations for us to calculate the permittivity including the self-interaction are the following:
{\bf 1}. The gauge invariance is a embarrassing and important problem.
Without considering the self-interaction, the $SU(N_c)$ gauge symmetry will be violated.
The color permittivity reflecting the characteristic of QGP must include the influence of self-interaction.
{\bf 2}.The non-Abelian permittivity contributed by the self-interaction will influence the relevant physical quantities of QGP. 
For example, the rough numerical result (\ref{final}) indicates that the nonlinear eigenfrequency shift is changed greatly by
this effect\refr{stinko}
$$
\Delta\omega ^S=-\0{\varp ^S}
{
\0{\pa \varp^L (\omega ,\vk )}{\pa \omega }|_{\omega _\vk}
}=0.282g^2T.$$

\end{document}